\documentclass[12pt]{article}
\usepackage{graphicx}
\usepackage{xcolor}
\definecolor{lcolor}{rgb}{0.5,0,0}
\definecolor{citcolor}{rgb}{0,0.3,0.0}
\usepackage[breaklinks,colorlinks,urlcolor=blue,citecolor=citcolor,linkcolor=lcolor]{hyperref}
\usepackage{amsmath}
\usepackage{tikz}

\newcommand\pubdate{\today}

\def\Title#1{\begin{center} {\Large #1 } \end{center}}
\def\Author#1{\begin{center}{ \sc #1} \end{center}}
\def\Address#1{\begin{center}{ \it #1} \end{center}}

\newcommand\pubblock{\rightline{\begin{tabular}{l}  \\ 
         \pubdate  \end{tabular}}}
\newenvironment{Abstract}{\begin{quotation}  }{\end{quotation}}
\newenvironment{Presented}{\begin{quotation} \begin{center} 
             PRESENTED AT\end{center}\bigskip 
      \begin{center}\begin{large}}{\end{large}\end{center} \end{quotation}}

\begin{document}
\begin{titlepage}
 \pubblock
\vfill
\Title{Proton structure functions in the dipole picture at next-to-leading order}
\vfill
\Author{H. Hänninen}
\Address{Department of Mathematics and Statistics, University of Jyväskylä,  P.O. Box 35, 40014 University of Jyväskylä, Finland}
\Author{H. Mäntysaari}
\Address{Department of Physics, University of Jyväskylä,  P.O. Box 35, 40014 University of Jyväskylä, Finland, and \\
Helsinki Institute of Physics, P.O. Box 64, 00014 University of Helsinki, Finland
}
\Author{R. Paatelainen}
\Address{Department of Physics and
Helsinki Institute of Physics, P.O. Box 64, 00014 University of Helsinki, Finland
}
\Author{J. Penttala}
\Address{Department of Physics, University of Jyväskylä,  P.O. Box 35, 40014 University of Jyväskylä, Finland, and \\
Helsinki Institute of Physics, P.O. Box 64, 00014 University of Helsinki, Finland
}
\vfill
\begin{Abstract}
 We predict heavy quark production cross sections in Deep Inelastic Scattering at high energy by applying the CGC effective theory. We demonstrate that when the calculation is performed consistently at next-to-leading order accuracy with massive quarks it becomes possible, for the first time in the dipole picture with perturbatively calculated center-of-mass energy evolution, to simultaneously describe both light and heavy quark production data at small $x$. We furthermore show how the heavy quark cross section data provides additional strong constraints on the extracted non-perturbative initial condition for the small-$x$ evolution equations.
\end{Abstract}
\vfill
\begin{Presented}
DIS2023: XXX International Workshop on Deep-Inelastic Scattering and
Related Subjects, \\
Michigan State University, USA, 27-31 March 2023 \\
     \includegraphics[width=9cm]{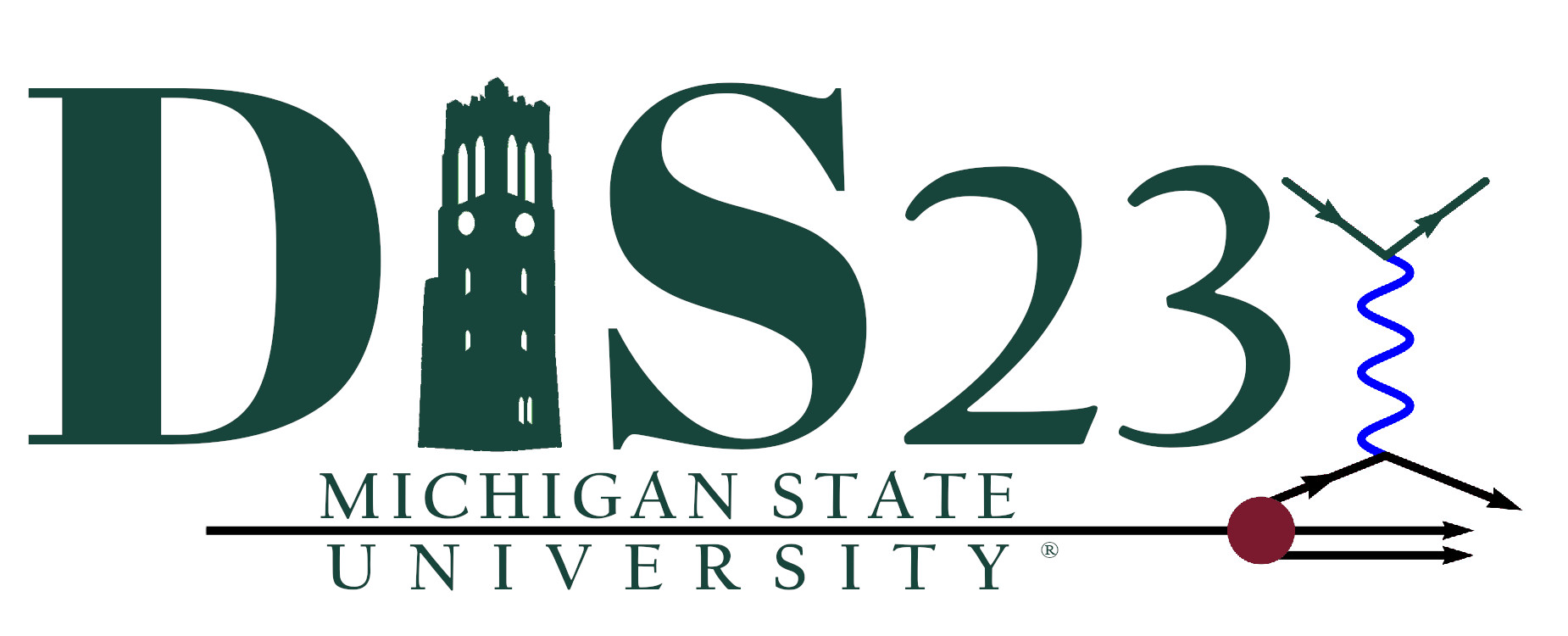}
     
\end{Presented}
\vfill
\end{titlepage}

\section{Introduction}
Total cross sections in Deep Inelastic Scattering are clean probes to access the internal structure of protons and nuclei. The precise structure function measurements from HERA~\cite{H1:2015ubc,H1:2018flt} have revealed that the gluon densities in the proton grow rapidly towards small longitudinal momentum fraction $x$. At high enough densities this growth should be limited by non-linear QCD dynamics. Accessing this saturation domain is one of the major physics goals of the future Electron-Ion Collider~\cite{AbdulKhalek:2021gbh} where it will also be possible to measure for the first time nuclear structure functions at small $x$. This is especially intriguing, as the higher density in the nucleus is expected to enhance the non-linear dynamics.

The precise structure function data is thus a stringent test for models including e.g. gluon saturation phenomena described in the Color Glass Condensate (CGC) effective theory of high-energy QCD~\cite{Gelis:2010nm}. Additionally, the DIS data is typically used to constrain the non-perturbative initial condition for the perturbative evolution equations. In the CGC framework, these are the BK and JIMWLK equations describing the energy or Bjorken-$x$ dependence of the hadron structure.

The leading order calculations in the CGC framework including perturbative small-$x$ evolution have been successful in describing the total cross section as measured at HERA, but it has not been possible to simultaneously describe both the total and charm production data~\cite{Albacete:2010sy,Lappi:2013zma}. In Ref.~\cite{Hanninen:2022gje} we demonstrated that when the calculation is consistently taken to next-to-leading order (NLO) accuracy a simultaneous describtion of all HERA structure function data in the small-$x$ region can be obtained.

\section{Structure functions at next-to-leading order}
In the dipole picture of DIS, valid at high energy, the incoming virtual photon fluctuates into a partonic Fock state long before the interaction with the target color field. This $q\bar q$, $q\bar q g$ or a higher order  state then scatters off the target, and the total cross section can be obtained from the forward elastic scattering amplitude by applying the optical theorem.

At NLO one has to consider both the $q\bar q$ contribution with loop corrections and the tree-level $q\bar q g$ contribution. The total virtual photon-target cross sections can then be written as
\begin{align}
    \sigma_{q\bar q}^{\gamma^*} &= K_{q\bar q} \otimes N_{01} \\
    \sigma_{q\bar q g}^{\gamma^*} &= K_{q\bar qg} \otimes N_{012}.
\end{align}
These two contributions are separately divergent, but the divergences cancel in their sum.
The perturbative impact factors $K_{q\bar q}$ and $K_{q\bar q g}$ are obtained in Ref.~\cite{Beuf:2021srj}. The $q\bar q$-proton and $q\bar q g$-proton interactions are described by $N_{01}$ and $N_{012}$ that satisfy the BK evolution equation. The BK equation is known at NLO accuracy~\cite{Balitsky:2008zza}, but due to itse computationally demanding form in Ref.~\cite{Hanninen:2022gje} we use approximative forms that include a resummation of most important transverse logarithms to all orders. These evolution equations are referred to as KCBK, ResumBK and TBK, see Refs.~\cite{Hanninen:2022gje,Beuf:2020dxl} for details. 

The precise HERA data is reported as reduced cross section defined as
\begin{equation}
    \sigma_r(y,x,Q^2) = F_2\left(x,Q^2\right) - \frac{y^2}{1+(1-y)^2} F_L\left(x,Q^2\right),
\end{equation}
where $y=Q^2/(sx)$ is the inelasticity, $\sqrt{s}$ is the center-of-mass energy for the electron-proton scattering and the structure functions $F_{2,L}$ are
\begin{align}
    F_2& = \frac{Q^2}{4\pi^2 \alpha_\mathrm{em}} \left(\sigma_T^{\gamma^*} +  \sigma_L^{\gamma^*}\right) \\
    F_L &= \frac{Q^2}{4\pi^2 \alpha_\mathrm{em}} \sigma_L^{\gamma^*}.
\end{align}
Here $T$ and $L$ refer to the virtual photon polarization.

\section{Results} 
Before the heavy quark contribution to the virtual photon wave function was available, the initial condition for the (approximative) NLO evolution was fitted to the constructed HERA light quark data in Ref.~\cite{Beuf:2020dxl}. In total 12 different fits were reported that differ by the choice of the resummation scheme, the running coupling scheme and the initial rapidity for the BK evolution.

We use all these 12 fits to calculate charm quark production at small $x<0.01$. We calculate the charm quark contribution to $\sigma_r$ and consider a fit to be allowed by the HERA data~\cite{H1:2018flt} if we obtain $\chi_c^2/N < 2.5$ with an optimal charm quark mass varied within $1.1\,\mathrm{GeV} < m_c < 1.6\,\mathrm{GeV}$. As a result, we find that only 3 out of the 12 fits are allowed by this data. The details of these fits are summarized in Ref.~\cite{Hanninen:2022gje}.

\begin{figure}[tb]
\begin{minipage}{.45\textwidth}
\centering
        \includegraphics[width=\textwidth]{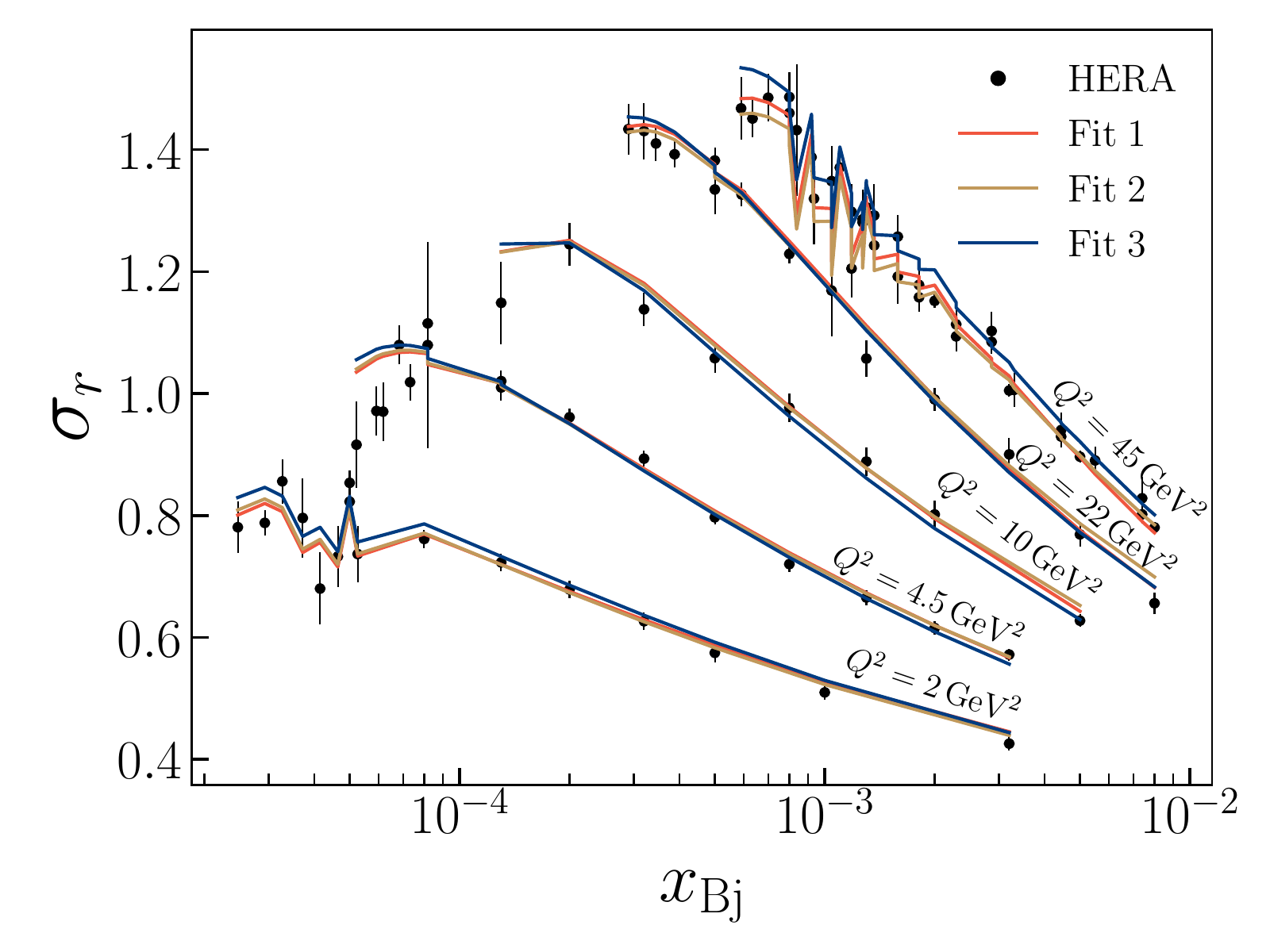} 
            \caption{Total reduced cross section compared to results obtained using the three fits allowed by the heavy quark data. Figure from Ref.~\cite{Hanninen:2022gje}, data~\cite{H1:2015ubc}. }
            \label{fig:inclusive}

\end{minipage}
\quad
\begin{minipage}{.45\textwidth}
\centering
\includegraphics[width=\textwidth]{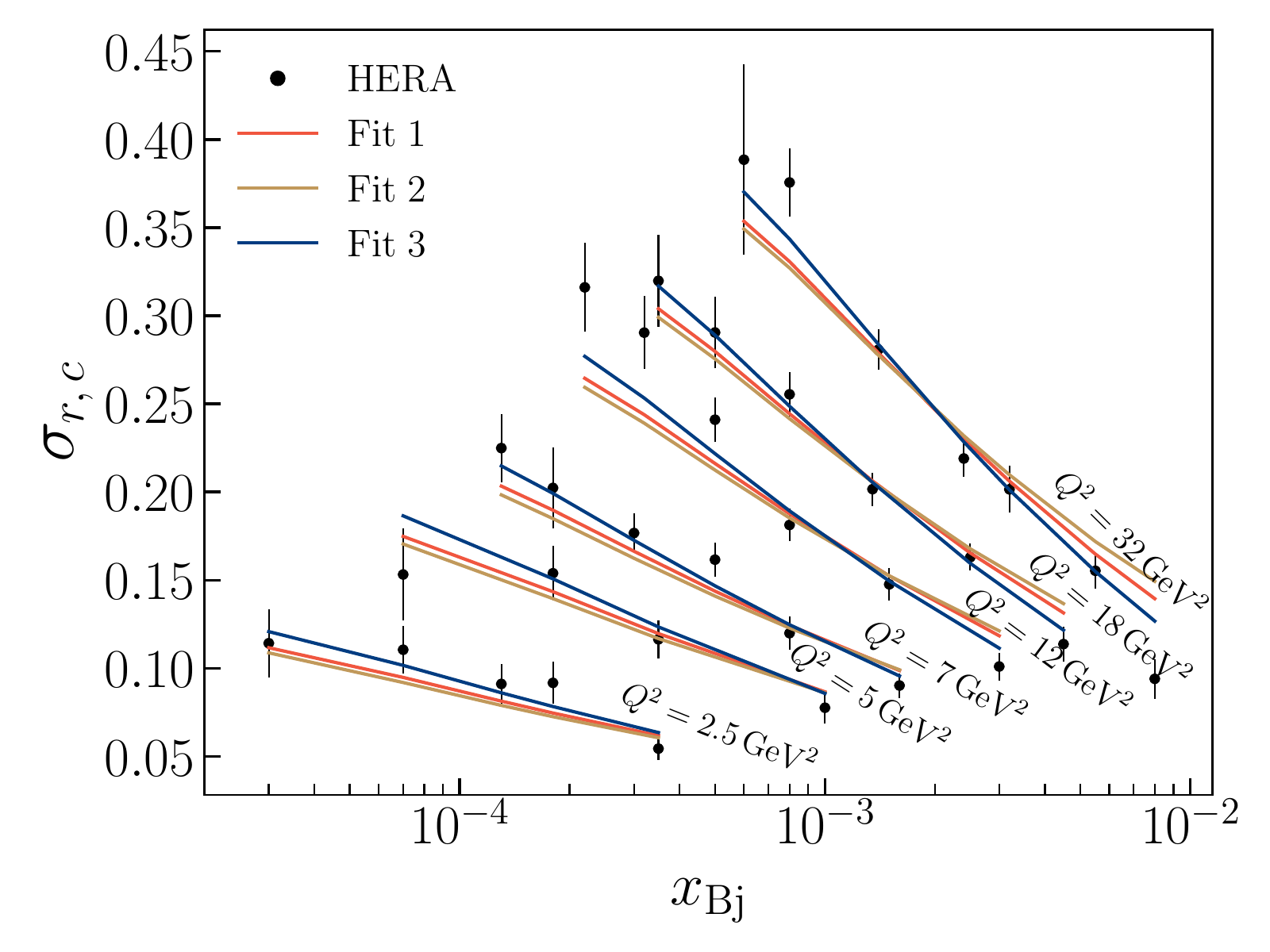} 
				\caption{Charm contribution to the reduced cross section obtained using the three fits allowed by the heavy quark data. Figure from Ref.~\cite{Hanninen:2022gje}, data~\cite{H1:2018flt}. }  
			\label{fig:charm}
\end{minipage}
\end{figure}

Using these three fits with the determined optical values for the charm quark mass we then calculate the reduced cross section and the charm contribution to it. The results are shown in Figs.~\ref{fig:inclusive}  and~\ref{fig:charm} in a few different $Q^2$ bins as a function of $x_\mathrm{Bj}$. Overall we find a very good description of all HERA small-$x$ structure function data, and emphasize that this is the first time when such a good agreement is obtained in the CGC framework including the perturbative BK evolution. We have also checked that these three fits are compatible with the $b$ quark production data~\cite{H1:2018flt}. Additionally, the fact that only a subset of the fits reported in~\cite{Beuf:2020dxl} are compatible with the heavy quark data demonstrates that these additional datasets will provide further constraints in the future global analyses. This is because the heavy quark production is sensitive to shorter distance scales (smaller dipoles) than the inclusive cross section.

\begin{figure}[h!]
\begin{minipage}{.48\textwidth}
\centering
        \includegraphics[width=\textwidth]{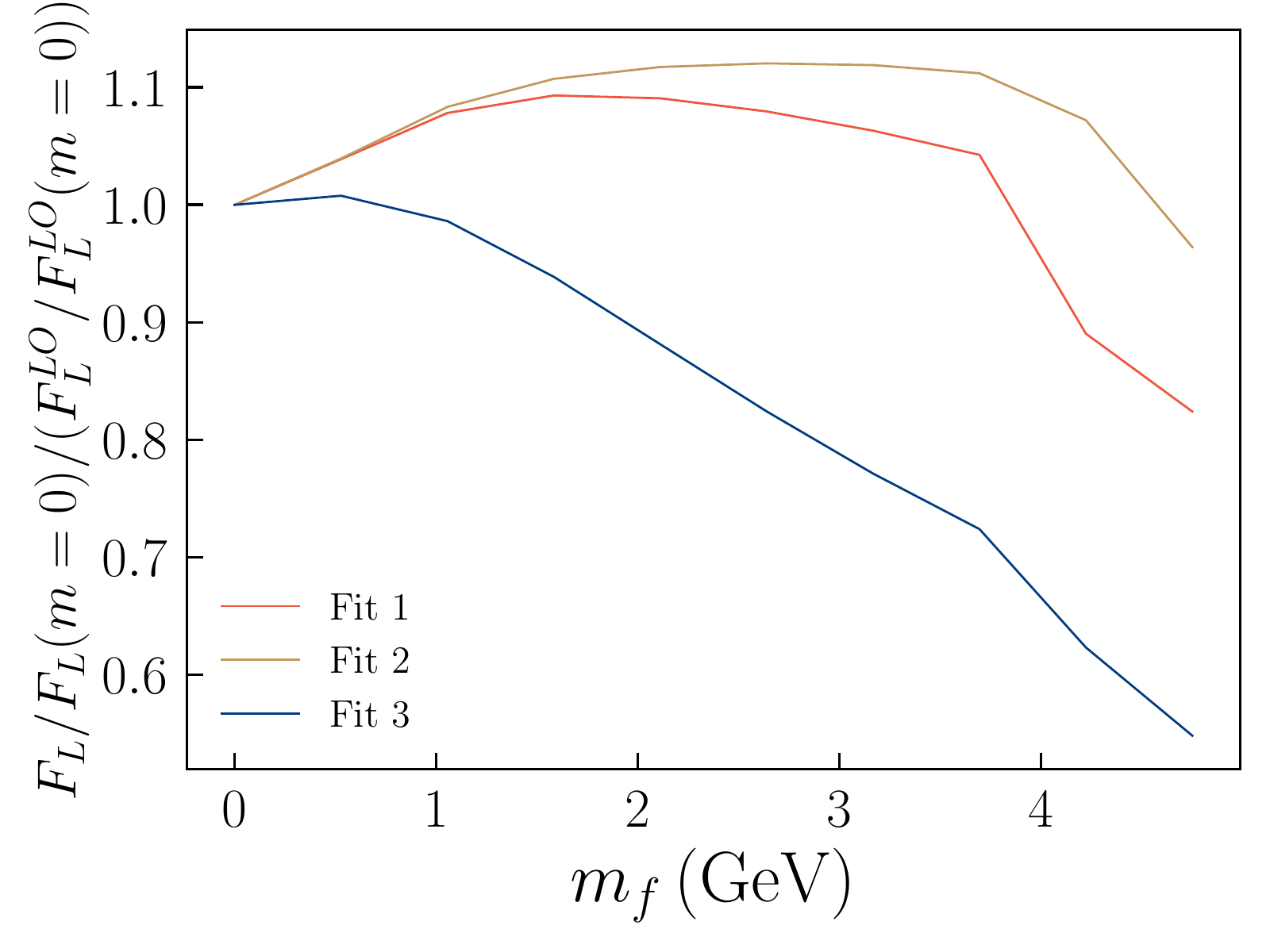} 
            \caption{Quark mass effect at NLO compared to LO in the case of longitudinal structure function. When the double ratio is above unity, the NLO impact factor enhances heavy quark production.  }
    \label{fig:nloimpfacmass}
\end{minipage}
\quad
\begin{minipage}{.48\textwidth}
\centering
\includegraphics[width=\textwidth]{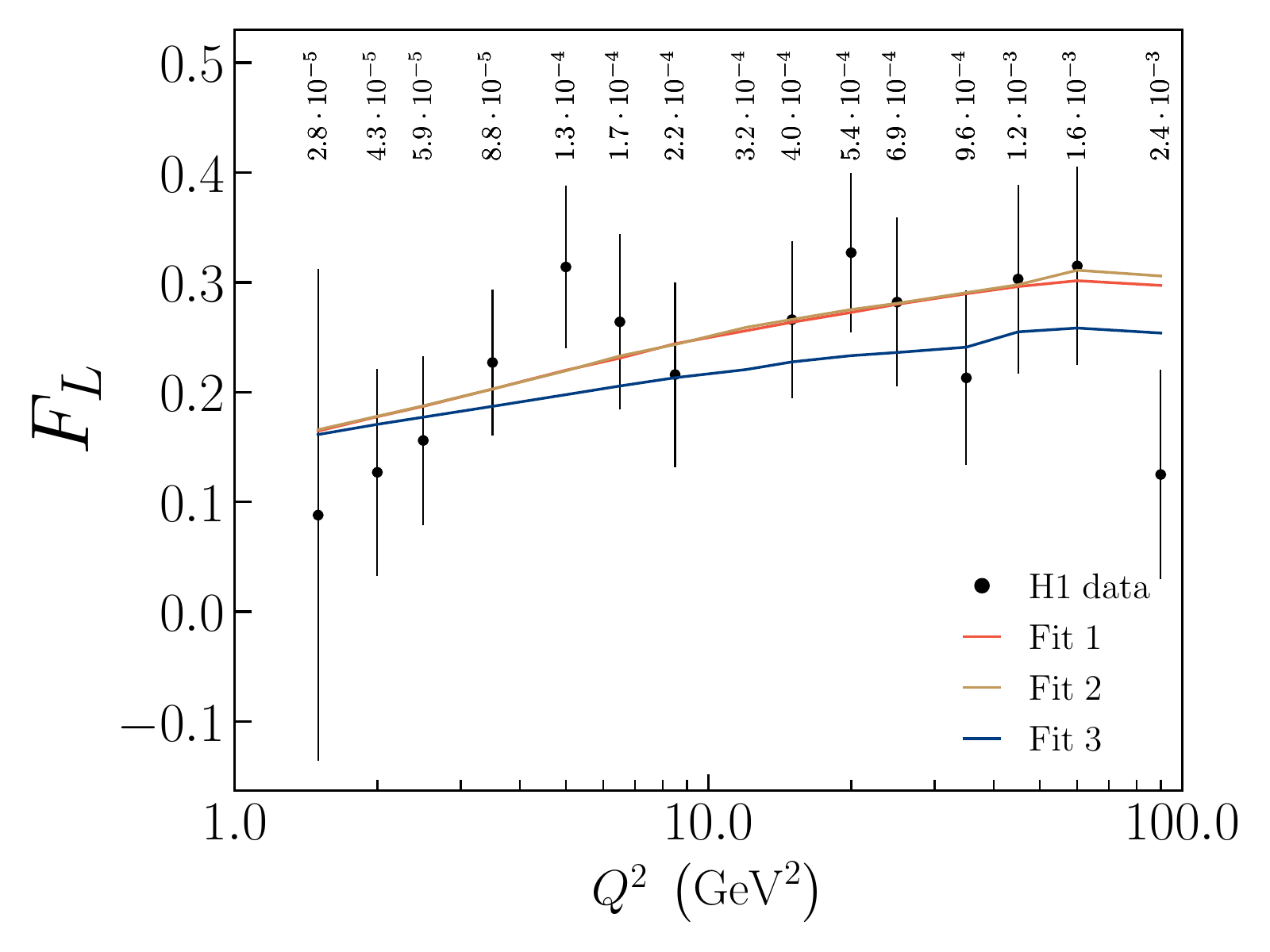} 
				\caption{Longitudinal structure function calculated using the three allowed NLO fits  compared to the HERA data~\cite{H1:2013ktq}. }  
			\label{fig:herafl}
\end{minipage}
\end{figure}

The NLO accuracy modifies the mass dependence of the cross section for two reasons. First, the NLO BK evolution does not develop a small anomalous dimension $\gamma \sim 0.8$ similarly as the leading order equation~\cite{Lappi:2016fmu}, which suppresses heavy quark production where small dipoles contribute. On the other hand, the NLO corrections to the impact factor enhance the heavy quark production when the ResumBK and KCBK evolution equations are used (the systematics is opposite with the TBK evolution). This is illustrated in Fig.~\ref{fig:nloimpfacmass} where the ratio of mass effects on the $F_L$ structure function at NLO and at LO is shown (using the same NLO dipole amplitude in NLO and LO calculation to isolate the effect of the impact factor). The net effect of these NLO corrections to both impact factor and to the evolution is such that the mass dependence at NLO matches that of the HERA data.

Finally we show comparison to the HERA $F_L$ data~\cite{H1:2013ktq} in Fig.~\ref{fig:herafl}. This data is not directly used in the fits. All three fits are found to be compatible with the HERA data, but future more precise $F_L$ measurements from the HERA can be expected to provide additional constraints. 

\section{Conclusions}
We have presented the first calculation of both inclusive and heavy quark DIS structure functions in the CGC framework which is compatible with the HERA small-$x$ structure function data. We have demonstrated that it is crucial to promote both the impact factor and the small-$x$ evolution equation to NLO accuracy in order to make global analyses feasible in the CGC framework. In addition to future global analyses with more flexible parametrizations for the initial condition of the small-$x$ evolution, we also plan to calculate precise predictions for the nuclear structure functions that will be measured at the EIC.

\emph{Acknowledgments}
This work was supported by the Academy of Finland, the Centre of Excellence in Quark Matter, the Centre of Excellence of Inverse Modelling and Imagining, and projects 338263, 346567, 321840, 347499 and 353772. This work was also supported under the European Union’s Horizon 2020 research and innovation programme by the European Research Council (ERC, grant agreement No. ERC-2018-ADG-835105 YoctoLHC) and by the STRONG-2020 project (grant agreement No. 824093).
J.P. is supported by the Finnish Cultural Foundation.
The content of this article does not reflect the official opinion of the European Union and responsibility for the information and views expressed therein lies entirely with the authors. 

\bibliographystyle{JHEP-2modlong.bst}
\bibliography{refs}
 
\end{document}